\documentstyle[twoside]{adacta}
\begin{document}
\headings{1}{6}
\newcommand{\pp}[1]{\phantom{#1}}
\newcommand{\ve}{\varepsilon}
\newcommand{\vs}{\varsigma}
\newcommand{\T}{{\,\rm Tr\,}}
\newcommand{\bbox}[1]{{\bf #1}}
\newcommand{\pol}{\frac{1}{2}}
\newcommand{\be}{\begin{equation}}
\newcommand{\ee}{\end{equation}}
\newcommand{\bear}{\begin{eqnarray}}
\newcommand{\eear}{\end{eqnarray}}
\newtheorem{proposition}{Proposition}
\newtheorem{fact}{Fact}
        \def\autor{M. Kuna}
        \def\nazov{Entanglement and pseudomixtures}
        \prvastrana=1
        \poslednastrana=6  
        \title{ENTANGLEMENT AND PSEUDOMIXTURES}
        \author{Maciej Kuna}
{Wydzia{\l} Fizyki Technicznej i Matematyki Stosowanej\\ 
Politechnika Gda\'{n}ska,
ul. Narutowicza 11/12, 80-952 Gda\'{n}sk, Poland}

\abstract{
In a recent paper Sanpera {\it et al.} have shown \cite{S}, that for
the simplest binary composite systems any density matrix can be
described in terms of only product vectors. The purpose of this
note  is to show that posibillity of decomposing any state as
pseudomixtures does not depend on dimension of the subsystems.
}

In a composite quantum system one can find states that involve
different degrees of correlation between subsystems. 
From the
physical point of view the most important characterization of
such states is related to the nature of these correlations.
{\it Separable\/} states
involve correlations which can be explained by classical models
\cite {Wer}. The notion of inseparability or entanglement 
is related to specific relations existing only in pure
quantum systems. Recently a general characterization of
separable states was given in the language of vectors from natural
cone of the Tomita-Takesaki theory, and quantum origin of this
notion was explained \cite{WAM}.

Quantum character of entaglement plays a crucial role in quantum
communication \cite{ben}, cryptography \cite{eke}, and quantum
computation \cite{bar}. A state of a composite quantum system
is {\it inseparable\/} if it cannot by represented as a convex mixture
of tensor products of states of its subsystems ($\rho \neq
\sum_{\omega} q_{\omega} {\rho}_{\omega}^{(1)} \otimes
{\rho}_{\omega}^{(2)}$; otherwise the state is separable). This definition
is very hard to handle. Therefore another characterizations are needed.
Peres \cite{per} and the Horodeckis  \cite{hor} have obtained a
characterization of separable states for systems described by
Hilbert spaces with dimensions 2 $\times$ 2 or 2 $\times$ 3.
Sanpera {\it et al.\/} have shown that for a binary system of
dimensions 2 $\times$ 2 any state can be described as a linear
combination of separable states called pseudomixtures.  

In this paper we show that any state of any composite quantum
system described by Hilbert spaces with finite dimensions can be
given in a form of a pseudomixture. In order to show this we use
a simple
modification of the definition of a separable state \cite{phor}:
\begin{fact}
Any density matrix $\rho$ is separable iff it is of the form:
\be 
\rho = \sum_{\alpha} p_{\alpha} P_{\alpha}\label{2}
\ee
where $p_{\alpha} > 0$, $\sum_{\alpha} p_{\alpha} = 1$ and
$P_{\alpha}$ are projectors on simple-tensor vectors.
\end{fact}
Let us consider two physical systems described by Hilbert spaces
${\cal H}_1$ and ${\cal H}_2$. A composite quantum system is
described by the Hilbert space ${\cal H}_1 \otimes {\cal H}_2$. We
need some facts about self-adjoint Hilbert-Schmidt operators 
$(\rho = {\rho}^*$ and
$\rho\in B_{\rm HS}({\cal H}_1\otimes~{\cal H}_2~))$:
\begin{proposition}
For any self-adjoint Hilbert-Schmidt operator $\rho$ and any
simple-tensor basis $\{\mid e_i
\otimes f_j\rangle\}_{ij}$ there exists an orthogonal
decomposition of $\rho$ into 
a diagonal component $A$ and an off-diagonal component $H$:
\be
\rho  = A + H \,\,\,\,\,\,\,{\rm and }\,\,\,\,\,\,\,(A,H)_{\rm HS} =
0,\label{1}
\ee
where $(A,H)_{\rm HS} = \T AH$ is a scalar product in $B_{\rm
HS}({\cal H}_1\otimes 
{\cal H}_2 )$.
\end{proposition}
{\bf Proof}: 
Let $\rho = \sum_t p_t P_t $ be the spectral decoposition of $\rho$.
Then we have:
\bear
 \rho &=& \sum_t p_t P_t = \sum_t p_t \mid\sum_{ij} {\alpha}^t_{ij} e_i
\otimes f_j\rangle\langle \sum_{kl} {\alpha}^t_{kl} e_k \otimes
f_l \mid \nonumber \\ 
 &=& \sum_{ijklt} p_t {\alpha}^t_{ij} \overline{{\alpha}^t_{kl}}  \mid e_i
 \otimes f_j\rangle\langle e_k \otimes f_l \mid \nonumber \\
 &=& \sum_{i,j,t} p_t \mid{\alpha}^t_{ij}{\mid}^2 \mid e_i
 \otimes f_j\rangle\langle e_i \otimes f_j \mid 
   + \sum_{i\neq k, j\neq l, t}
 p_t {\alpha}^t_{ij} \overline{{\alpha}^t_{kl}}  \mid e_i
 \otimes f_j\rangle\langle e_k \otimes f_l \mid \nonumber \\ 
 &\equiv&  A + H
\eear
 and
\bear
 AH&=&\sum_{m,n,{t}} p_{t}
\mid{\alpha}^{t}_{mn}{\mid}^2 \mid {e_m} \otimes
{f_n}\rangle\langle {e_m}\otimes {f_n} 
 \mid 
 \sum_{i\neq k, j\neq l, {s}}
 p_{s} {\alpha}^{s}_{ij} \overline{{\alpha}^{s}_{kl}}  \mid {e_i}
 \otimes {f_j}\rangle\langle {e_k} \otimes {f_l} \mid \nonumber \\ 
 &=& \sum_{m,n,{t}} p_{t}\mid{\alpha}^{t}_{mn}{\mid}^2 
 \sum_{i\neq k, j\neq l, {s}}
 p_{s} {\alpha}^{s}_{ij} \overline{{\alpha}^{s}_{kl}} 
 \mid {e_m} \otimes {f_n}\rangle\langle {e_m}\otimes {f_n}  \mid \mid {e_i}
 \otimes {f_j}\rangle\langle {e_k} \otimes {f_l} \mid \nonumber \\ 
 &=&\sum_{i\neq k, j\neq l, t,s}
 p_{t}p_{s}\mid{\alpha}^{t}_{ij}{\mid}^2  {\alpha}^{s}_{ij}
 \overline{{\alpha}^{s}_{kl}}  
 \mid {e_i} \otimes {f_j}\rangle\langle {e_k} \otimes {f_l} \mid .
\eear
Therefore $\T AH = 0$.\hfill$\Box$
 
\begin{proposition}
If for any decomposition (\ref{1}) all diagonal components $A$ are equal
to 0, then $\rho$ is equal to 0.
\end{proposition}
{\bf Proof}:
If we have a decomposition (\ref{1}) associated with a basis $\{\mid e_i
\otimes f_j\rangle\}_{ij}$, then any other decomposition can
be obtained by a
unitary transformations $ e_i = \sum_{k} u_{ik} \hat{e}_{k}$ and
$f_j = \sum_{l} v_{jl}\hat{f}_{l}$ in the following way:
\bear
 P_t &=& \mid\sum_{ij} {\alpha}^t_{ij} e_i \otimes f_j\rangle
 \langle \sum_{mn} {\alpha}^t_{mn} e_m \otimes f_n\mid\nonumber\\
 &=&  \mid\sum_{ij} {\alpha}^t_{ij} (\sum_{k_1} u_{ik_1} \hat{e}_{k_1})
 \otimes (\sum_{l_1} v_{jl_1}\hat{f}_{l_1})\rangle
\langle \sum_{mn} {\alpha}^t_{mn} (\sum_{k_2} u_{mk_2}
\hat{e}_{k_2}) \otimes 
 (\sum_{l_2} v_{nl_2}\hat{f}_{l_2})\mid
 \nonumber \\
 &=& \mid\sum_{k_{1}l_{1}} (\sum_{ij}{\alpha}^t_{ij}
 u_{ik_1}v_{jl_1})\hat{e}_{k_1} \otimes \hat{f}_{l_1}\rangle
 \langle \sum_{k_{2}l_{2}} (\sum_{mn}{\alpha}^t_{mn}
 u_{mk_2}v_{nl_2})\hat{e}_{k_2} \otimes \hat{f}_{l_2}\mid
 \nonumber \\
 &=& \sum_{k_{1}l_{1}k_{2}k_{1}l_{2}} \sum_{ijmn}{\alpha}^t_{ij} 
 \overline{{\alpha}^t_{mn}} 
 u_{ik_1}v_{jl_1}\overline{u_{mk_2}}\overline{v_{nl_2}}\mid
 \hat{e}_{k_1} \otimes \hat{f}_{l_1}\rangle
 \langle \hat{e}_{k_2} \otimes \hat{f}_{l_2}\mid .
\eear

It follows that the operators with the decomposition associated with
the basis \break $\{\mid \hat{e}_{k}\otimes
\hat{f}_{l}\rangle\}_{kl}$ (which will be 
described by $\hat A$ and $\hat H$) can be given as a function of
${\alpha}^t_{ij}$ (these coefficients will be referred to as 
the coefficients  of spectral projectors in the basis  $\{\mid e_i
\otimes f_j\rangle\}_{ij}$):
\be
 \hat{A} = \sum_t p_t\sum_{k,l} \sum_{ijmn}{\alpha}^t_{ij} 
 \overline{{\alpha}^t_{mn}} 
 u_{ik}v_{jl}\overline{u_{mk}}\overline{v_{nl}}\mid
 \hat{e}_{k} \otimes \hat{f}_{l}\rangle
 \langle \hat{e}_{k} \otimes \hat{f}_{l}\mid
\ee
and
\be
 \hat{H} = \sum_t p_t \sum_{k_{1},l_{1}, k_{2}\neq k_{1}, l_{2} \neq l_{1}
 } \sum_{ijmn}{\alpha}^t_{ij} 
 \overline{{\alpha}^t_{mn}}
 u_{ik_1}v_{jl_1}\overline{u_{mk_2}}\overline{v_{nl_2}}\mid
 \hat{e}_{k_1} \otimes \hat{f}_{l_1}\rangle
 \langle \hat{e}_{k_2} \otimes \hat{f}_{l_2}\mid .
\ee
Putting ${\beta}_{ijmn} = \sum_t p_t {\alpha}^t_{ij} 
\overline{{\alpha}^t_{mn}}$ and using Hermiticity of $\rho$ we obtain:
\be
 \hat{A} = \sum_{k,l} \sum_{i \geq m,j \geq n} 2 \Re ({\beta}_{ijmn}
 u_{ik}v_{jl}\overline{u_{mk}}\overline{v_{nl}})\mid
 \hat{e}_{k} \otimes \hat{f}_{l}\rangle
 \langle \hat{e}_{k} \otimes \hat{f}_{l}\mid
\ee
and
\be
 \hat{H} = \sum_{k_{1},l_{1},k_{2}\neq k_{1}, l_{2} \neq l_{1}} 
 \sum_{i \geq m,j \geq n} 2 \Re ({\beta}_{ijmn}
 u_{ik_1}v_{jl_1}\overline{u_{mk_2}}\overline{v_{nl_2}})\mid
 \hat{e}_{k_1} \otimes \hat{f}_{l_1}\rangle
 \langle \hat{e}_{k_2} \otimes \hat{f}_{l_2}\mid .
 \ee
Assume that all operators $A = 0$ for any decomposition (\ref{1}). It
means that
\be 
 \forall_{u \in U_1 , v \in U_2 }\forall_{kl}\sum_{i > m,j > n} 2 \Re
 ({\beta}_{ijmn} u_{ik}v_{jl}\overline{u_{mk}}\overline{v_{nl}}) =
 0,
\ee
where $U_i$ is the set of all unitary transformations on ${\cal H}_i$.
We can choose unitary transformations which in the $k_{o}$-th column has 
only two nonzero elements $u_{ak_{o}} =
\frac{1}{\sqrt{2}} ({\delta}_{ai_{o}} + {\delta}_{am_{o}})$ and
analogously $v_{bl_{o}}
 =\frac{1}{\sqrt{2}} ({\delta}_{bj_{o}} + {\delta}_{bn_{o}})$.
For such $u$ and $v$ we have:
\be
 \frac{1}{2}\Re ({\beta}_{i_o j_o m_o n_o }) = 0.
\ee
Next, choosing $u_{ak_{o}} =\frac{1}{\sqrt{2}} (i{\delta}_{ai_{o}} +
{\delta}_{am_{o}})$ and $v$ as before, we have
\be
 \frac{1}{2}\Re (i{\beta}_{i_o j_o m_o n_o })= 
 \frac{1}{2}\Im ({\beta}_{i_o j_o m_o n_o })= 0.
\ee
Then ${\beta}_{i_o j_o m_o n_o }= 0$ for all $i_o \neq m_o$ and $j_o \neq
n_o$, so $\rho = 0$.\hfill$\Box$
 
\begin{proposition}
If $\rho$ is a density matrix then a norm of the operator $ H$ is less
than $\frac{1}{2}$  and its trace is equal to 0.
\end{proposition}
{\bf Proof}: 
We can write the operator $H$ in the following form:
\be
 H =  \sum_{i\neq k, j\neq l, t}
 p_t {\alpha}^t_{ij} \overline{{\alpha}^t_{kl}}  \mid e_i
\otimes f_j\rangle\langle e_k \otimes f_l \mid = \sum_{i>k, j>l,
t} p_t G^t_{ijkl} . 
\ee
The operator $G^t_{ijkl}$ has two eigenvalues $\mid{\alpha}^t_{ij}
{\alpha}^t_{kl}\mid$ and $-\mid{\alpha}^t_{ij}{\alpha}^t_{kl}\mid$.
Since
\be
 0\leq(\mid{\alpha}^t_{ij}\mid - \mid{\alpha}^t_{kl}\mid)^2 =
 \mid{\alpha}^t_{ij}{\mid}^2 + \mid{\alpha}^t_{kl}{\mid}^2
 -2\mid{\alpha}^t_{ij}{\alpha}^t_{kl}\mid ,
\ee 
then
\be
 \mid{\alpha}^t_{ij}{\alpha}^t_{kl}\mid \leq {1\over
 2}(\mid{\alpha}^t_{ij}{\mid}^2 + \mid{\alpha}^t_{kl}{\mid}^2) .
\ee
Using this we can estimate the norm of the operator $H$:
\bear
 {\parallel} H{\parallel}&=&{\parallel}\sum_{i>k, j>l, t} p_t
 G^t_{ijkl}{\parallel}\leq 
 \sum_{i>k, j>l, t} p_t {\parallel} G^t_{ijkl}{\parallel} 
 = \sum_{i>k, j>l, t} p_t \mid{\alpha}^t_{ij}{\alpha}^t_{kl}\mid 
 \nonumber \\ 
 &\leq& \sum_{i>k, j>l, t} p_t {1\over 2}(\mid{\alpha}^t_{ij}{\mid}^2 +
 \mid{\alpha}^t_{kl}{\mid}^2) 
 \leq {1\over 4}\sum_{t} p_t \sum_{ijkl}
 (\mid{\alpha}^t_{ij}{\mid}^2 + \mid{\alpha}^t_{kl}{\mid}^2) =
 {1\over 2} .
\eear 

The last equality results from $\sum_t
p_t = 1$ and $\sum_{ij}\mid{\alpha}^t_{ij}{\mid}^2 = 1$.
Trace of this operator equals 0:
\bear
 1 &=& \T \rho = \T (\sum_{i,j,t} p_t
 \mid{\alpha}^t_{ij}{\mid}^2 \mid e_i
 \otimes f_j\rangle\langle e_i \otimes f_j \mid ) + \T H\nonumber \\
  &=& \sum_{i,j,t} p_t
 \mid{\alpha}^t_{ij}{\mid}^2 \T (\mid e_i
 \otimes f_j\rangle\langle e_i \otimes f_j \mid ) + \T H
 = \sum_{i,j,t} p_t
 \mid{\alpha}^t_{ij}{\mid}^2 + \T H\nonumber \\
 &=& \sum_{t} p_t \sum_{i,j}
 \mid{\alpha}^t_{ij}{\mid}^2  + \T H = 1 + \T H .
\eear\hfill$\Box$
 
Let $\rho$ be any density matrix from  $B_{\rm HS}({\cal
H}_1\otimes {\cal H}_2)$. For 
this matrix we can choose such a decomposition (\ref{1}) that $\T
\hat{A}^2$ has a maximal value. This maximum exists because we take
supremum over $U_1\otimes U_2$, so over a compact set, of a continuous 
function of $u$ and $v$:
\be
 \T \hat{A}^2 = \sum_{k,l} (\sum_t p_t \mid \sum_{ij}{\alpha}^t_{ij}  
 u_{ik}v_{jl}{\mid}^2 )^2 ,
\ee
Therefore there exists a basis realising this maximum, namely
$\{\mid {e_i^0} \otimes {f_j^0}\rangle\}_{ij}$:
\bear
 \rho &=& \sum_{i=k,j=l,{t_{o}}} p_{t_{o}} \mid{\beta}^{t_{o}}_{ij}{\mid}^2
 \mid {e_i^0}
 \otimes {f_j^0}\rangle\langle {e_i^0} \otimes {f_j^0} \mid
 + \sum_{i\neq k, j\neq l, {t_{o}}}
 p_{t_{o}} {\beta}^{t_{o}}_{ij} \overline{{\beta}^{t_{o}}_{kl}}  \mid {e_i^0}
 \otimes {f_j^0}\rangle\langle {e_k^0} \otimes {f_l^0} \mid \nonumber \\ 
&\equiv&  A(1) + H(1) \,\,\,\,\,\,{\rm and} \,\,\,\,\,\, \T
A(1)^2 = \sup_{A \in {\cal A}} \T A^2   ,
\eear
where ${\cal A}$ describe the set of all diagonal
components in simple-tensor decomposition,  $p_{t_{o}}$  and
${\beta}^{t_{o}}_{ij}$  are the eigenvalues of $\rho$ 
and the coefficients of spectral
projectors in a basis $\{\mid {e_i^0} \otimes {f_j^0}\rangle\}_{ij}$,
respectively. 

From Proposition 3 we know that ${\parallel} H(1){\parallel} < 
\frac{1}{2}$. The operator $A(1)$ is a linear (in the first step
even convex) combination of simple tensors, so we can think
about this decomposition as
separating out the simple-tensor part. 
Since $H(1)$ is self-adjoint (as a sum of self-adjoint operators
$G^t_{ijkl}$) we can continue the separation procedure.
We again choose a decomposition with maximum $\T A^2$ which is associated
with the basis $\{\mid {e_i^1} \otimes {f_j^1}\rangle\}_{ij}$ :
\bear
 H(1) &=& \sum_{s(1)} o_{s(1)} O_{s(1)} 
 = \sum_{i,j,{s(1)}} o_{s(1)}
 \mid{\beta}^{s(1)}_{ij}{\mid}^2 \mid e_i^1 
 \otimes f_j^1\rangle\langle e_i^1 \otimes f_j^1 \mid \nonumber \\ 
 &+& \sum_{i\neq k, j\neq l, {s(1)}}
 o_{s(1)} {\beta}^{s(1)}_{ij} \overline{{\beta}^{s(1)}_{kl}}  \mid e_i^1
 \otimes f_j^1\rangle\langle e_k^1 \otimes f_l^1 \mid 
 = A(2) + H(2) 
\eear
After $n$ steps we obtain:
\be 
 \rho = \sum_{i=1}^{n} A(i)  +  H(n)
\ee
Since the decomposition is orthogonal, we have
\be 
 \T H^{2}(n - 1) = \T A^{2}(n)  + \T H^{2}(n) 
\ee
and
\be 
 \sum_{s(n - 1)}\mid o_{s(n - 1)}{\mid}^2 =  \sum_{i,j}(\sum_{s(n)}
 o_{s(n)} \mid{\beta}^{s(n)}_{ij}{\mid}^2 )^2 +
 \sum_{s(n)}\mid o_{s(n)}{\mid}^2 > \sum_{s(n)}\mid o_{s(n)}{\mid}^2
\ee
The equality $\T H^2 (n - 1) = \T H^2 (n)$ can be obtained iff $A(n) =0$
and, therefore, by Proposition 2 only for $H(n - 1) = 0$.
Then  a sum of squares of eigenvalues of $H(n)$ is less then a sum of
squares of eigenvalues of $H(n - 1)$.  This means that $\lim_n
{\parallel}H(n){\parallel} = 0$.
In the limit of these procedures we obtain a mixture of simple-tensor
projectors \break $\{\mid {e_i^u}
\otimes {f_j^u}\rangle\langle {e_i^u} \otimes {f_j^u}\mid
\}_{iju}$, not necessarily 
orthogonal to one another, with real
coefficients. If we collect positive and negative coefficients separetely
we obtain a pseudomixture:
\be
 \rho = a {\rho}^{sep(+)} - b {\rho}^{sep(-)} ,\label{3}
\ee
where $a,b \geq 0$, $a = 1 + b$, and ${\rho}^{sep(+)}$ and
${\rho}^{sep(-)}$ are 
separable states. For the simplest binary composite systems,
it was shown \cite{S} that by using geometrical properties typical of
this dimension 
it is possible to determine cardinality of ${\rho}^{sep(+)}$ and
${\rho}^{sep(-)}$ i.e. the smallest $\alpha$ in the decomposition
(\ref{2}). Our procedure does not give such an information. The
decomposition (\ref{3}), similarly to other decompositions of
this kind,  is not
unique. The most interesting pseudomixture is the one for
which $b$ is minimal. Our procedure does not determine such a
pseudomixture. 

Since in our procedure we use only the properties of the
Hilbert-Schmidt scalar product,
it seems that it can be
transferred also to composite systems described by infite-dimensional
Hilbert spaces. 
The language of vectors from natural
cone of the Tomita-Takesaki theory is the most natural one for such a
problem. The infinite-dimensional case will be discussed elsewhere
\cite{wamk}. 

\medskip 
I would like to thank for the hospitality to Centrum Leo Apostel
in Brussels where a part of this work was done. I am indebted to
W.A.Majewski and M.Czachor for comments. The work was supported
by the Polish-Flemish Grant No. 007.

\end{document}